\documentclass[preprint,showpacs,prl,amsmath,floatfix]{revtex4}
\usepackage{graphicx}
\usepackage{epsfig}
\usepackage{amssymb}
\begin{document}
\title{Force induced conformational transition in a system of 
interacting stiff polymer: Application to unfolding}

\author{Sanjay Kumar$^\dagger$ and Debaprasad Giri$^\ddagger$} 
\affiliation{Department of Physics, Banaras Hindu University,
Varanasi 221 005, India \\
$^\dagger$ Max-Planck-Institute for the Physics of Complex Systems, 
Noethnitzer, Strabe 38, 01187, Dresden, Germany \\
$^\ddagger$Physics Section, MMV, Banaras Hindu University,
Varanasi 221 005, India.}
\email{SK: yashankit@yahoo.com; DG: dgiri@bhu.ac.in}
\date{\today}

\begin{abstract}
We consider a stiff polymer chain in poor solvent and apply 
a force at one end of the chain. We find that by varying 
the stiffness parameter, polymer undergoes a transition 
from the globule state to the folded like  state. The conformation
of folded state mimics the $\beta$-sheet as seen in titin molecule. 
Using exact enumeration technique, we study the extension-force 
and force-temperature diagrams of such a system. Force-temperature 
diagram shows the re-entrance behaviour for flexible chain. 
However, for stiff chain this re-entrance behaviour is absent 
and there is an enhancement in $\theta$-temperature with the 
rise of stiffness. We further propose that the internal 
information about the frozen structure of polymer can be read 
from the distribution of end-to-end distance which shows 
saw-tooth like behaviour. 
\end{abstract}

\maketitle

Conformational and structural properties of bio-polymers play an 
important role in the biological phenomena. Protein folding is a 
phenomenon which is associated with the primary structure of the 
chain. Their stability or resistance to unfolding are investigated 
either by varying pH of the solution or by varying the temperature 
\cite{1}. Many proteins unfold due to the force instead of change in 
the chemical environment or temperature. For example, the effect 
of stress on the titin molecule is a force induced transition from 
the native state to the unfolded state \cite{2}. The presence of 
strong hysteresis together with sudden jump seen in the force
 {\em vs} extension curve suggests that the unfolding is a first 
order transition \cite{3}. 
Theoretically, this type of transition has been studied in the 
framework of statistical mechanics considering a flexible polymer 
chain in a poor solvent under the influence of external force 
\cite{4}, and for DNA type polymers under an unzipping force 
\cite{5}. However, stiffness plays an important role as is seen
in worm like chain (WLC) model \cite{2,3}, which has 
not been incorporated in the lattice model so far to explain the 
phase diagram of such molecules under the influence of the external force.

Recently the phase diagram of homo semi-flexible polymer chain 
with zero force has been studied by mean field theory \cite{6} 
and Monte Carlo simulation \cite{7} which shows three distinct phases  
namely : (i) an open coil phase at high temperature, 
(ii) a molten globule collapse at low temperature and 
low stiffness, and (iii) a `frozen' or `folded' state at low 
temperature and large stiffness. Our aim in this paper is to 
describe the effect of external force applied at one end of the 
stiff polymer chain in poor solvent. We consider 
self-attracting self-avoiding walks (SAWs) on a square lattice 
\cite{8}. One end of the chain is subjected to an external force 
while the other end is kept fixed. The stretching energy $E_s$ 
arising due to the applied force $F$ is taken as 

\begin {equation}
E_s= -F. x
\end {equation}
where $x$ is the $x$-component of end-to-end distance 
$(\mid x_1-x_N \mid)$.  Stiffness in the chain is introduced 
by associating an energy barrier $\epsilon_b$ with every `turn' 
(or bend) of the walk \cite{9}.  We associate a negative energy 
$\epsilon_u$ for each non-bonded occupied nearest neighbour pairs. 
The partition function of such a system may be written as
\begin{equation}
Z_N  = \sum_{(N_b, N_u, \mid x \mid)} C_N (N_b, N_u, \mid x \mid) 
b^{N_b} u^{N_u} \omega^{\mid x \mid}
\end{equation}

$C_N (N_b, N_u, \mid x \mid)$ is the total number of SAWs \cite{10} 
of $N$ steps having $N_b$ turns (bends)  and $N_u$ nearest neighbour 
pairs respectively.  $\omega$ is the Boltzmann weight for the 
force which is defined as $\exp[\beta (F.\hat{x})]$, where $\hat{x}$
is the unit vector along the {\em x}-axis. $\beta$ is defined as 
$\frac{1}{k T}$ where $k$ is the Boltzmann constant and T is the 
temperature. $b = \exp (-\beta \epsilon_b)$ and $u = \exp (-\beta 
\epsilon_u)$ are Boltzmann weights of bending and nearest neighbour 
interaction respectively. 

\begin{figure}
\centerline{\includegraphics[width=3in]{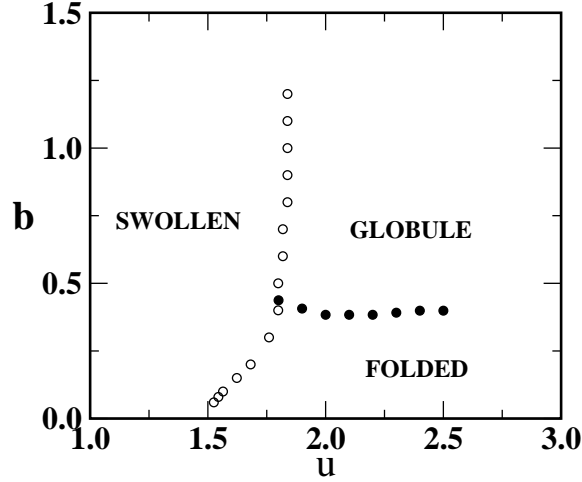}}
\caption{Phase diagram shows the variation of $b$ (Boltzmann
weight of bending) with $u$ (Boltzmann weight of nearest
neighbour interaction) of a linear semi-flexible polymer chain
in two dimensional space. The collapsed state includes both
globule and folded state. The open circle represents transition
from swollen state to the collapsed state and the filled circle
represents transition from the folded state to the globule state.}
\label{fig-1}
\end{figure}

Since the polymer is on the lattice where continuous overall 
rotations are not possible, we assume that under the applied force 
there is an alignment of protein along the force direction with 
zero torque. This is equivalent to the assumption that the relaxation 
for the overall rotational degrees of freedom is much shorter than 
that of the structural relaxation that is responsible for unfolding 
or folding process. Apart from logarithmic factors, the low temperature 
limit of the rotationally averaged contribution to the the free energy 
due to tension obtained in the context of rubber elasticity \cite{11}
coincides with Eq. (1).

In all the single molecule experiments, a finite size of the chain 
is used and the fact that no true phase transition can occur in a 
single macromolecule, we calculate ``state diagram" instead of 
``phase diagram" \cite{12}. We obtained $C_N (N_b, N_u, \mid x \mid)$ 
up to $N \leq 30$ steps walk on a square lattice by the exact 
enumeration method and use it to calculate $Z_N (b, u, \omega)$.  
The boundaries in the state diagram may be found for $F=0$ from 
the maxima of fluctuation of $N_u (=\frac{\partial <N_u>}{\partial 
\epsilon_u}$) or fluctuation of $N_b (=\frac{\partial <N_b>}{\partial 
\epsilon_b})$. The sudden change in the non-bonded nearest neighbors  $N_u$ 
indicates a phase transition; hence maxima in the derivative of $N_u$ 
with respect to $\epsilon_u$ indicates the location of a phase 
transition.  The swollen to collapsed transition line (which is also 
called the $\theta$-line) is obtained from the peak value of 
$\frac{\partial<N_u>}{\partial \epsilon_u}$ for fixed value of $b$ 
(shown in Fig. 1 by open circle).  Folded to globule transition 
line (shown in Fig. 1 by filled circle) has been obtained from the 
peak value of $\frac{\partial<N_b>}{\partial\epsilon_b}$ at fixed $u$. 

There are three states marked by swollen, frozen (or folded) and 
globule as shown in Fig. 1. For $b=1$, we restore the value of 
$u_c = 1.93$ using extrapolation scheme \cite{13} for flexible 
polymer chain, that is in good agreement with the value found by 
Foster \cite{14} and the Monte Carlo simulation result ($1.94 \pm 
0.005$) \cite{15}. However, we shall confine ourselves here to the 
constant force ensemble to get the exact boundaries for 
finite $N$. 

\begin{figure}[b]
\centerline{\includegraphics[width=5in]{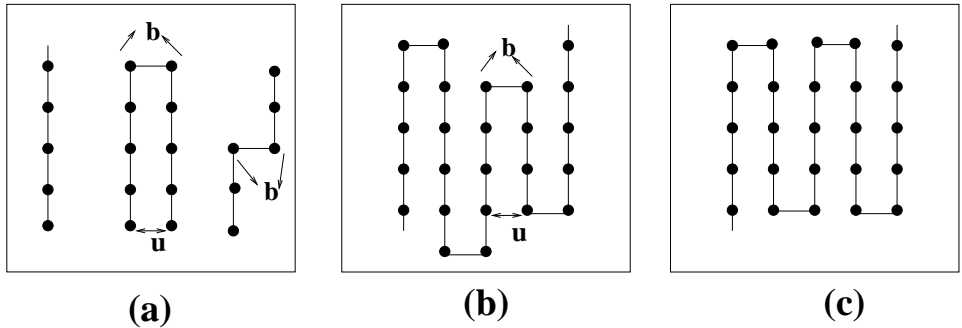}}
\caption{Typical conformations of polymer chain at higher
value of $u$ and lower value of $b$ in two dimensional
space: (a) This represents a situation where hairpin
structure will be preferred than the other conformations;
(b) The resulting conformation at large $N$;
(c) The conformation at $T = 0$. This can be mapped by 
Hamiltonian walk.}
\label{fig-3}
\end{figure}

From Fig. 1, we find that the $\theta$-line bends to left as stiffness 
increases ($b$ decreases) indicating the enhancement of 
$\theta$-temperature. Intuitively one might think that stiffness 
favors the extended state but the fact is that stiffness favors a 
folded state in a poor solvent at low temperature. It is obvious 
that on a square lattice hairpin like structures (Fig. 2(a)) 
minimize the 
number of folds and maximize the number of nearest neighbors. 
Due to this increase in nearest neighbours,
there is a decrease in $u$ ({\it i.e.} rise in $\theta$-temperature).
Such trend
in $\theta$-line has also been observed by extensive Monte Carlo
simulation by Bastolla and Grassberger \cite{7}.
For large $N$, resulting configurations will give rise frozen
or folded like state and the partition function 
will be then dominated by configurations similar to Fig. 2(b). 
Since the number of such configurations are very small, entropy 
associated with folded state will be quite low. 
The length of the fold will be of the  order of 
$N^{1/2}$ and depends on $(\beta \epsilon_b)$.

\begin{figure}[t]
\label{fig-2}
\centerline{\includegraphics[width=5in]{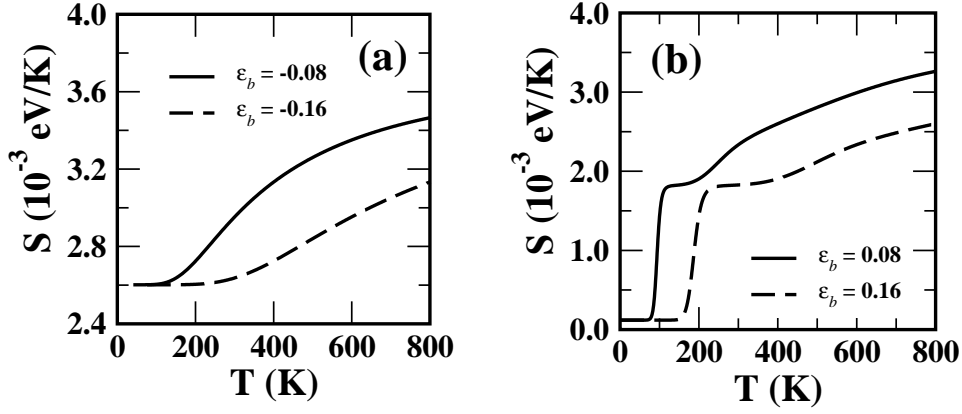}}
\caption{Plot of Entropy ($S$) {\em vs}  Temperature ($T$) 
for different values of stiffness parameter ($\epsilon_b$).} 
\end{figure}

To study the order of transition from the frozen state to the globule 
state at zero force, we calculated the entropy ($S$) using the 
following relation of free energy ($G$):
\begin{eqnarray}
G &=& -k T \ln Z_N (T) \\
S &=& - \left ( \frac{\partial G}{\partial T} \right ) 
\end{eqnarray}

We use two different sets of  $\epsilon_b$ at fixed $u=4.0$ and 
plotted the entropy with temperature in Fig. 3. 
In the first set (Fig. 3(a)), we choose $\epsilon_b 
= -0.08 \; {\rm eV}$ and $-0.16 \; {\rm eV}$, implying $b=\exp(+ 
\beta \epsilon_b)$ is greater than one, and this corresponds to 
a flexible chain.  For the second set (shown in Fig. 3(b)) $\epsilon_b$ 
is positive ($b < 1$), and this corresponds to the stiff chain. 
It can be seen that for the stiff chain there is a sudden jump in 
entropy (shown in Fig. 3(b)) corresponding to a transition 
from the folded state to the globule state while it is absent in 
the case of a flexible polymer chain.

Although in principle the frozen structures (Fig. 2(b)) can be seen
everywhere in the ``Folded" state, numerically it is easier to
observe this for smaller $b$ ($ < 0.2$) and larger $u$ ($ > 1.93$).
In this regime we have numerically computed the Boltzmann contribution 
of individual conformation of polymer chain in the partition function 
and verified that the maximal contribution comes from the structure
similar to Fig. 2(b) (like $\beta$-sheet seen in titin molecules).
Hence in this region, it will be possible to study
the effect of force on the unfolding transition by applying force
at one end of the chain. To do so
we set  $\epsilon_u = -1$ and Boltzmann constant $k = 1$ and 
study force-temperature and extension-force curve.  It can be 
shown that in this scale $F = (\frac{\ln \omega}{\ln u})$, 
$T = 1/\ln(u)$ and $\epsilon_b = -(\frac{\ln b}{\ln u})$.

The variation of critical force with temperature for unfolding
transition where polymer goes from the collapsed state to the 
unfolded state (or extended state) is shown in Fig. 4. 
Note that the collapsed state consists  of 
both the globule and the folded state. The transition line separates 
the collapsed state from the unfolded state. It is not possible to
see transition from the folded to the globule  
state because it is not induced by the force.
It is interesting to note that for a stiff polymer chain, 
critical force increases monotonically with temperature 
and becomes almost constant at very low temperature. 
However, in the case of flexible polymer chain, 
we observe a phenomenon of `re-entrance' 
{\it i.e.} the critical force goes through maximum as 
temperature is lowered. For example, one can see that the polymer chain at 
fixed force (say $F=0.9$), is in the extended state at low 
temperature.  With the increase in temperature, the chain 
is found in the collapsed state.  With the further rise of 
temperature, it acquires again the conformations of the 
extended (swollen) 
state. Similar re-entrance behaviour is also found in the 
transfer matrix calculation of the directed walk models of 
flexible polymer chain \cite{16}.  However, if we introduce 
stiffness in the chain, the re-entrance behaviour is found to be 
suppressed. This is because in the globule state, the entropy 
associated with the flexible chain is very high, while the 
stretched chain has almost zero entropy ({\it i.e} polymer
forms a rod like shape).
This indicates that polymer goes from 
a high entropy state to the low entropy state under the 
application of force for flexible polymer chain. However, 
in the case of stiff polymer chain, collapsed state has frozen 
structure with almost zero entropy, therefore, no re-entrance 
is observed in going from the frozen state to the extended state.
This can be seen using phenomenological argument near $T = 0$ 
where the conformation of polymer chain in the collapsed state
looks like Fig. 2(c) similar to the conformations formed by the 
Hamiltonian walks. 
Thus, from the principle of balance of energy, the free
energy of folded state  and the free energy of 
stretched state due to the force (using Eq. 1) can be equated 
as \cite{13} 
\begin{equation}
-F N =  N \epsilon_u + 2 \sqrt{N}(\epsilon_b - \epsilon_u) - N T S_c
\end {equation}

The second term of right hand side in Eq. (5) is a surface correction 
term which also includes the bending energy. The third term is 
due to the entropy associated with the collapsed state ($S_c$ per 
monomer).  
Minimization of energy with respect to $N$ and substituting 
$\epsilon_u = -1$, Eq. (5) gives  
\begin{equation}
F_c(T)=1-\frac{(1+\epsilon_b)}{\sqrt{N}}+T S_c
\end{equation}

\begin{figure}[t]
\centerline{\includegraphics[width=2.8in]{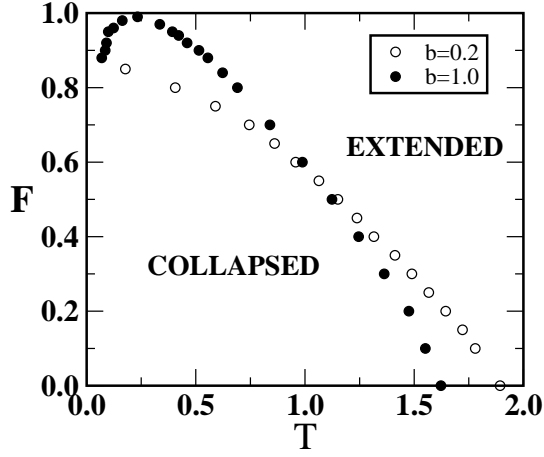}}
\caption{Force-Temperature phase diagram for fixed $b$.
Here collapsed state consists of both the globule and the
folded state.}
\label{fig-4}
\end{figure}

For large $N$, the second term goes to zero. For finite $N=30$ (as
we have taken in our study), 
the correction term is important which gives $F_c = 0.8174$. 
This is in good  agreement with the value found from 
Fig. 4 near $T=0$.  It is to be noted that for a flexible 
polymer, entropy associated with globule is finite and hence 
there is a positive slope ($\frac{dF_c}{dT}$), while for 
stiff chain entropy 
associated with frozen structure is almost zero and hence
there is no slope, therefore, we do not find any re-entrance 
behavior in this case.

\begin{figure}
\centerline{\includegraphics[width=2.8in]{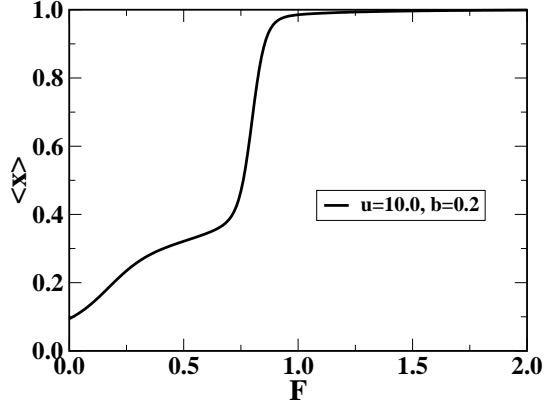}}
\caption{ Plot of $<x>$ vs $F$ for fixed $u$ and $b$,
Here $u=10.0$ and $b=0.2$. }
\label{fig-5}
\end{figure}

The extension {\it vs} force ($<x>\; vs \; F$) curve is 
shown in Fig. 5. 
From the figure, it is evident that for small forces, a polymer chain
is in the compact folded state and slightly oriented along 
the force direction.  At larger forces, the polymer chain 
has the conformations similar to  the extended (swollen) structure. 
Completely stretched states can be obtained only by applying
very high forces. A similar behaviour is also 
seen in the case of flexible polymer chain under the influence 
of external forces \cite{17}. Molecular dynamics simulation
of protein under the stress of an external denaturating force 
acting on a terminal end or on the entire chain \cite{18} shows the 
intermediate stages during the unfolding and the extension-force
curves are similar to Fig. 5 found by us.

We also study the probability distribution curves $P(x)$ 
with $x$ for flexible and stiff chains defined by
\begin{equation}
P(\mid x \mid)= \frac{1}{Z_N (b, u, \omega)}\sum_{(\mid x \mid)}
 C_N (N_b, N_u, \mid x \mid) b^{N_b} u^{N_u} \omega^{\mid x \mid}.
\end{equation}
In Fig. 6, we have shown $P(x)$ for 
different values of $\omega$ and at fixed $u=3.0$ corresponding to
the collapsed state. 

The probability distribution curves have many interesting 
features. For flexible chain {\it i.e.}  $b=1$, the maxima 
of $P(x)$ corresponds to the collapsed state at $F=0$ 
($\omega = 1.0$). For higher force, $\omega = 20$  both 
flexible and semi-flexible polymers are  found to be in 
the ``rod-like" state. However, at intermediate force, 
the probability distribution curve has ``saw-tooth" type 
of behaviour for the stiff chain while it is continuous for 
the flexible chain. The $x$-component of end-to-end 
distribution function gives information about the 
internal structure of the folded state by applying the 
suitable force. For small forces, the thermal fluctuations 
are too weak to unfold the polymer chain, and  it stays
in the folded state most of the time. This fact is more 
or less reflected in the structureless distribution 
function with a well defined peak at the most likely value 
of the end-to-end distance. In contrast, when we apply a 
force close to the critical force, the small thermal 
fluctuations are sufficient to open up the loops in the 
$\beta$-sheet (as shown in Fig. 2(b)). Here, one sees the 
statistical unfolding events in the structure of the 
distribution functions in the form of the ``saw-tooth'' 
kind of behaviour for the stiff chain while it is continuous 
for the flexible chain. The presence of intermediate stages 
seen during the unfolding in the molecular dynamic simulation 
\cite{18} and the one found by us in the  probability distribution 
curve, suggests the experimentalist to "tease" the folded polymer 
(proteins) by some external force and then measure the distribution 
function. This may reveal some interesting information about 
the folded state. 

\begin{figure}
\centerline{\includegraphics[width=6in]{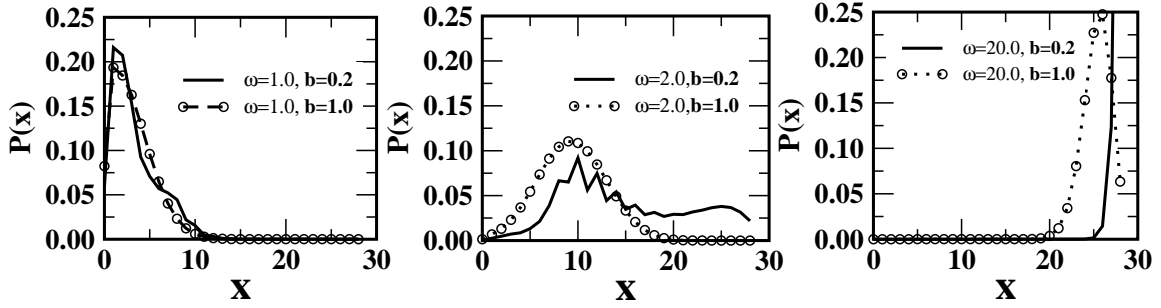}}
\caption{Plot of $P(x)$ vs $x$ for flexible ($b=1.0$) 
and stiff ($b=0.2)$) polymer chain at different 
$\omega$ (Boltzmann weight of the force). 
At intermediate force the saw-tooth behaviour in the 
probability distribution gives signature of unfolding 
events in the frozen structure.}
\label{fig-6}
\end{figure}

In this paper we have studied the complete state diagram 
of stiff polymer chain under the influence of external 
force. We showed the existence of the folded-like state 
in stiff homopolymer. We have also found that there is 
an enhancement of $\theta$-temperature (decrease in $u$) 
with the rise of stiffness parameter $\epsilon_b$. 
The absence of re-entrance in the stiff chain has been 
explained by using the phenomenological argument. We have also 
tried to explain the unfolding event as seen in titin molecule
on the basis of the probability distribution curve.

\acknowledgments

We thank Y. Singh for many helpful discussions on the 
subject and Erwin Frey for suggesting us to study the 
extension-force curve of stiff polymer through series 
analysis. Financial assistance from INSA, New Delhi 
and DST, New Delhi are acknowledged.


\begin{thebibliography}{99}
\bibitem{10} The total number of conformations ($C_{30}$) are 
16741957935348 and total CPU time is  arround 45 days on 
Pentium-IV 1.7 GHz machine.
\end{thebibliography}
\end{document}